# MIMO Relaying Broadcast Channels With Linear Precoding and Quantized Channel State Information Feedback

Wei Xu, *Member, IEEE*, Xiaodai Dong, *Senior Member, IEEE*, and Wu-Sheng Lu, *Fellow, IEEE*

*Abstract*—Multi-antenna relaying has emerged as a promising technology to enhance the system performance in cellular networks. However, when precoding techniques are utilized to obtain multi-antenna gains, the system generally requires channel state information (CSI) at the transmitters. We consider a linear precoding scheme in a MIMO relaying broadcast channel with quantized CSI feedback from both two-hop links. With this scheme, each remote user feeds back its quantized CSI to the relay, and the relay sends back the quantized precoding information to the base station (BS). An upper bound on the rate loss due to quantized channel knowledge is first characterized. Then, in order to maintain the rate loss within a predetermined gap for growing SNRs, a strategy of scaling quantization quality of both two-hop links is proposed. It is revealed that the numbers of feedback bits of both links should scale linearly with the transmit power at the relay, while only the bit number of feedback from the relay to the BS needs to grow with the increasing transmit power at the BS. Numerical results are provided to verify the proposed strategy for feedback quality control.

## I. INTRODUCTION

RELAYING technologies have attracted considerable attentions due to its promising potential of improving the system capacity and increasing the coverage at a low cost in next generation wireless communication systems including 3GPP LTE-Advanced and IEEE 802.16m [1]–[3]. Since multiple-input multiple-output (MIMO) has been a well-known technology for next generation wireless applications, recent studies of the subject matter have focused on investigating different relaying strategies in multi-antenna systems.

For a simple three-node system with one source and one destination communicating through a relay, the design of amplify-and-forward/decode-and-forward (AF/DF) relaying has been well investigated [4]–[6]. The capacity bounds are studied in [4] for a regenerative MIMO relay system from an information-theoretic point of view. Concerning nonregenerative MIMO relays, the optimal precoding design are presented in [5] and [6] aiming at channel capacity maximization. Theoretic capacity scaling law is analyzed in [7] for systems with multiple relays. In [8], a relay selection strategy is proposed for achieving additional multi-relay diversity gain for practical applications. These studies have been focused on systems serving a remote terminal via relay nodes.

With multiple antennas deployed for increasing multiplexing gain, works have been under way to investigate point to multipoint (multiple terminals) systems which are usually referred to as multiuser broadcast channels (BCs) or multiple access channels (MACs). For a MIMO BC/MAC without relay, the theoretic capacity limits have been investigated in the literature [9], [10], and the dirty paper coding (DPC) is known as a capacity-achieving technique for a MIMO BC [11]. On the other hand, because of the high complexity of DPC implementation, several efficient linear precoding techniques have been proposed, see, e.g., [12] and [13]. Another recent trend of research interest is the design of AF relay in multiuser MIMO systems. Especially in cellular networks, employing fixed relay stations has been found efficient in dealing with users located in a so-called "dead zone" [1], [14]. For the MIMO relay BC, the problem of joint precoding design has been studied in [14]–[16], while the MIMO MAC with relaying has been examined in [17].

A common technical drawback in the above relaying schemes is that they all require full channel state information (CSI) of both two hop channels at all transmitter sides, and thus can be infeasible for many practical wireless applications, particularly in frequency-division duplexed (FDD) systems. A common way to deal with this problem in MIMO channels is to exploit finite-rate feedback, in that each receiver node feeds back a finite number of bits of its CSI at the beginning of each transmission block. This technique has been widely considered for both point-to-point MIMO [18], [19] and multiuser MIMO systems [20]–[25]. So far as we know, however, little work has considered *quantized* CSI feedback schemes in precoded MIMO relay systems, with [26] and [27] as notable and interesting exceptions. The studies in [26] propose a limited feedback scheme for systems using Grassmannian beamforming in a three-node two hop system with an AF relay,

Manuscript received April 28, 2010; accepted June 19, 2010. Date of publication July 08, 2010; date of current version September 15, 2010. The associate editor coordinating the review of this manuscript and approving it for publication was Prof. Ye (Geoffrey) Li. This work was done when W. Xu was with the Department of Electrical and Computer Engineering, University of Victoria. This work was supported in part by the Natural Sciences and Engineering Research Council of Canada through Grant 349722-07. Part of this work was presented at the IEEE International Symposium on Information Theory (ISIT), Austin, TX, June 2010.

W. Xu was with the Department of Electrical and Computer Engineering, University of Victoria, Victoria V8W 3P6, Canada. He is now with the National Mobile Communications Research Laboratory (NCRL) and the School of Information Science and Engineering, Southeast University, Nanjing 210096, China (e-mail: wxu.seu@gmail.com).

X. Dong and W.-S. Lu are with the Department of Electrical and Computer Engineering, University of Victoria, Victoria V8W 3P6, Canada (e-mail: {xdong@ece.uvic.ca; wslu@ece.uvic.ca}).

Color versions of one or more of the figures in this paper are available online at http://ieeexplore.ieee.org.

Digital Object Identifier 10.1109/TSP.2010.2056687

while [27] extends the limited feedback design for precoding strategies serving multiple data streams simultaneously. Unlike the point-to-point two-hop systems, the design of precoding schemes in [26] and [27] is different and becomes more complex for point-to-multipoint relay broadcast systems even when perfect CSI are available [14]–[16]. Moreover, when considering precoding techniques with quantized CSI feedback, the system performance will be affected by the channel quantization quality of both two-hop channels. In this work, by assuming only quantized CSI is available at the transmitters through feedback, we study the multi-antenna downlink system where the base station (BS) broadcasts multiple data streams to a number of remote users via the help of a multi-antenna AF relay. The main contributions of this paper are summarized as follows.

- A linear precoding scheme for a MIMO relaying BC using quantized CSI feedback is proposed. In this scheme, the relay feeds back $B_1$ bits to the BS for determining the beamforming vector of each data stream, and each user terminal sends $B_2$ bits of its quantized CSI to the relay for calculating the downlink beamforming at the relay. We characterize the upper bounded rate achieved by the proposed precoding scheme at high SNRs, which is known as the interference-limited effect in limited feedback multiuser systems.
- In order to guarantee the rate loss relative to the precoding scheme with perfect CSI within a given constant, we develop a strategy of controlling the feedback quality, i.e., the numbers of quantization bits $B_1$ and $B_2$, for both two-hop channels. For limited feedback multiuser systems without relaying, study in [20] has revealed that the number of feedback bits per user should be increased linearly with the system SNR (in decibels) to maintain a constant rate loss. Our study extends the existing result in [20] to MIMO-relay assisted multiuser systems and show that the feedback quality grows with both the BS transmit power and the relay transmit power according to

$$\frac{B_1}{M-1} = \log_2 P_2 - \log_2\left(M + \frac{N}{P_1}\right) + \alpha \quad (1)$$

and

$$\frac{B_2}{N-1} = \log_2 P_2 + \alpha \quad (2)$$

where $M$ and $N$ are the numbers of antennas at the BS and relay, respectively, $P_1$ and $P_2$ are the transmit power constraints at the BS and the relay, respectively, and $\alpha$ is an offset for the feedback quality control which depends on the required rate loss.

The rest of the paper is organized as follows. Section II gives a brief description of the MIMO relaying broadcast system under consideration. Section III presents a linear precoding scheme using quantized CSI feedback from the receivers. In Section IV, the achievable rate by the proposed scheme is analyzed and a rate upper bound due to quantized feedback is characterized for high SNR cases. Then, in order to guarantee a maximal rate loss in the system, a quantization quality control strategy is proposed in Section V. Concluding remarks are made in Section VI.

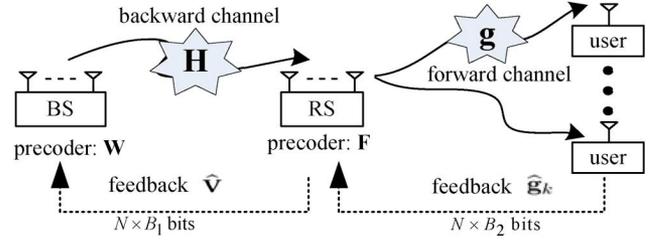

Fig. 1. System model of a relay-assisted multi-antenna downlink channel.

## II. SYSTEM MODEL

Consider a multiuser downlink channel with an $M$-antenna BS serving $K$ single-antenna users through an $N$-antenna relay station (RS). The system model is depicted in Fig. 1. For simplicity, we assume that the BS has more than or at least equal number of antennas as in the RS, that is $M \geq N$. To focus our study on the effect of the limited feedback scheme, we consider the system with $K = N$. Note that for $K > N$, the system will first implement user scheduling before transmission [14].

Let $\mathbf{x}_s \in \mathbb{C}^{N \times 1}$ be the transmit symbol vector at the source. Before transmitting it to the relay, the source pre-processes $\mathbf{x}_s$ by an $M \times N$ precoding matrix $\mathbf{W}$, the received symbol vector at the RS can thus be written by

$$\mathbf{r} = \sqrt{\rho_1}\mathbf{H}\mathbf{W}\mathbf{x}_s + \mathbf{n} \quad (3)$$

where $\mathbf{H} \in \mathbb{C}^{N \times M}$ is the backward channel matrix from the BS to the RS, $\mathbf{n}$ is the complex Gaussian noise with zero-mean and unity variance, and $\rho_1$ is a factor to scale the transmit symbol such that the total power constraint $P_1$ at the BS is satisfied. The energy of the input symbols $\mathbf{x}_s$ is normalized by $\mathbb{E}[\mathbf{x}_s\mathbf{x}_s^H] = \mathbf{I}$ where $\mathbb{E}[\cdot]$ represents the expectation operator. It follows that the BS transmit power constraint is given by

$$\mathrm{tr}(\mathbf{W}\mathbf{W}^H) = \frac{P_1}{\rho_1}. \quad (4)$$

Assume that the backward channel obeys the Rayleigh distribution and the value of $P_1$ has taken into account the path loss effect of the channel. The entries of the channel matrix $\mathbf{H}$ can be treated as independent complex Gaussian random variables with zero-mean and unity variance. Here, the direct links between the source and remote users have been neglected due to large path loss and severe shadowing effects.

After receiving the symbols, the relay pre-processes the received vector by an $N \times N$ precoding matrix $\mathbf{F}$, then it broadcasts the precoded symbols to distributed users. Let $\rho_2$ be a scaling factor for the transmit symbol at the relay. The RS forwards the symbol

$$\mathbf{x}_r = \sqrt{\rho_1\rho_2}\mathbf{F}\mathbf{H}\mathbf{W}\mathbf{x}_s + \sqrt{\rho_2}\mathbf{F}\mathbf{n} \quad (5)$$

with the relay transmit power constraint $P_2$ formulated as

$$\mathbb{E}_{\mathbf{H}}\left[\rho_1\mathrm{tr}(\mathbf{F}\mathbf{H}\mathbf{W}\mathbf{W}^H\mathbf{H}^H\mathbf{F}^H) + \mathrm{tr}(\mathbf{F}\mathbf{F}^H)\right] = \frac{P_2}{\rho_2} \quad (6)$$

which, like in [8], [25], is an average transmit power constraint at the RS. Under these circumstances, the received symbol at user $k$ is

$$y_k = \sqrt{\rho_1\rho_2}\mathbf{g}_k^H\mathbf{FHW}\mathbf{x}_s + \sqrt{\rho_2}\mathbf{g}_k^H\mathbf{Fn} + z_k \quad (7)$$

where $\mathbf{g}_k^H$ is the $1 \times N$ channel vector between the RS and the $k$th user terminal, and $z_k$ is the complex Gaussian noise at the $k$th user terminal with zero mean and unity variance. Assume that $\mathbf{g}_k$ is also Rayleigh distributed and its entries are zero mean complex Gaussian variables with unity variance. Without loss of generality, we define the $k$th entry $x_k$ in the transmit vector $\mathbf{x}_s$ as the desired information for user $k$. By representing the BS precoding matrix as $\mathbf{W} = [\mathbf{w}_1 \cdots \mathbf{w}_N]$, where $\mathbf{w}_k$ is the beamformer for data $x_k$, the SINR at user $k$ can be calculated by

$$\gamma_k = \frac{\left|\mathbf{g}_k^H\mathbf{FHw}_k\right|^2}{\sum_{j\neq k}\left|\mathbf{g}_k^H\mathbf{FHw}_j\right|^2 + \frac{1}{\rho_1}\left\|\mathbf{g}_k^H\mathbf{F}\right\|^2 + \frac{1}{\rho_1\rho_2}}. \quad (8)$$

Accordingly, the averaged achievable sum rate of the system is given by

$$R = \frac{1}{2}\mathbb{E}_{\mathbf{H},\mathbf{G}}\left[\sum_{k=1}^{N}\log(1+\gamma_k)\right] \quad (9)$$

where $\mathbf{G} = [\mathbf{g}_1 \cdots \mathbf{g}_N]^H$ represents the concatenation of relay-to-user channels and the factor $1/2$ results from the fact that data is transmitted over two time-slots.

## III. LINEAR PRECODING WITH QUANTIZED CSI FEEDBACK

In a MIMO relay BC, it is difficult to obtain the optimal design of $\mathbf{W}$ and $\mathbf{F}$ for joint source and relay precoding. Although some suboptimal joint precoding optimization methods have been proposed in [14]–[16], they are in general based on iterative mechanisms and require full CSI of both two-hop channels at all transmitters. As a result, these methods are computationally complex and they cannot be directly applied in our system due to the lack of perfect CSI. In this study, we exploit the structured source and relay precoding scheme proposed in the previous work [15] and extend it to scenarios using quantized CSI feedback.

### A. Linear Precoding Design

In this subsection, we briefly review the structured precoding scheme proposed in [15] where the source precoding exploits a singular value decomposition (SVD)-based precoding matrix and the relay precoding combines the SVD-based receiving matrix and the zero-forcing beamforming (ZFBF) scheme. Applying SVD to $\mathbf{H}$ under $N \leq M$, it gives

$$\mathbf{H} = \mathbf{U}[\boldsymbol{\Sigma}\ \mathbf{0}][\mathbf{V}\ \mathbf{V}_r]^H \quad (10)$$

where $\boldsymbol{\Sigma}$ is an $N \times N$ diagonal matrix and $\mathbf{V}$ is a column unitary matrix with size $M \times N$. The precoding matrices are then specified as

$$\mathbf{W} = \mathbf{V} \quad (11)$$

and

$$\mathbf{F} = \mathbf{F}_1\mathbf{U}^H \quad (12)$$

where $\mathbf{F}_1 = [\mathbf{f}_1 \cdots \mathbf{f}_N]$ with $\mathbf{f}_k$ being the normalized $k$th column of the matrix $\mathbf{G}^H(\mathbf{G}\mathbf{G}^H)^{-1}$. Notice that $\mathbf{F}_1$ follows the same design of conventional ZFBF in the multiuser BC [20], and this pair of precoding matrices diagonalize both the source-to-relay channel and the concatenated relay-to-user channel. By substituting the structured precoding matrices into (7), we have the stacked received symbols as

$$\begin{aligned}\mathbf{y} &= \sqrt{\rho_1\rho_2}\mathbf{G}\mathbf{F}_1\mathbf{U}^H\mathbf{H}\mathbf{V}\mathbf{x}_s + \sqrt{\rho_2}\mathbf{G}\mathbf{F}_1\mathbf{U}^H\mathbf{n} + \mathbf{z}\\ &= \sqrt{\rho_1\rho_2}\boldsymbol{\Xi}\boldsymbol{\Sigma}\mathbf{x}_s + \sqrt{\rho_2}\boldsymbol{\Xi}\tilde{\mathbf{n}} + \mathbf{z}\end{aligned} \quad (13)$$

where $\mathbf{y} = [y_1 \cdots y_N]^T$ is the concatenated received symbols at users, $\boldsymbol{\Xi}$ is a diagonal matrix with its diagonal elements as $\mathbf{g}_k^H\mathbf{f}_k$ ($k = 1,\cdots,N$), and $\tilde{\mathbf{n}} \in \mathbb{C}^{L\times 1}$ is the equivalent Gaussian noise with unit variance. From (13), when this structured precoding is utilized with perfect CSI, the SINR in (8) for receiver $k$ reduces to

$$\gamma_k^P = \frac{\left|\mathbf{g}_k^H\mathbf{F}_1\boldsymbol{\Sigma}\mathbf{V}^H\mathbf{v}_k\right|^2}{\frac{1}{\rho_1}\left\|\mathbf{g}_k^H\mathbf{F}_1\right\|^2 + \frac{1}{\rho_1\rho_2}} = \frac{\sigma_k^2\left|\mathbf{g}_k^H\mathbf{f}_k\right|^2}{\frac{1}{\rho_1}\left|\mathbf{g}_k^H\mathbf{f}_k\right|^2 + \frac{1}{\rho_1\rho_2}} \quad (14)$$

where $\sigma_k$ is the $k$th diagonal entry of $\boldsymbol{\Sigma}$ and $\mathbf{v}_k$ is the $k$th column of $\mathbf{V}$.

Given the structured precoding matrices, by considering the power constraints in (4) and (6), we can thus determine the two power scaling parameters as follows:

$$\rho_1 = \frac{P_1}{N} \quad (15)$$

and

$$\begin{aligned}\rho_2 &= \frac{P_2}{\text{tr}\left(\mathbf{F}_1^H\mathbf{F}_1\mathbb{E}_\mathbf{H}\left[\left(\frac{P_1}{N}\boldsymbol{\Sigma}^2 + \mathbf{I}\right)\right]\right)}\\ &= \frac{P_2}{\left(\frac{P_1M}{N}+1\right)\text{tr}\left(\mathbf{F}_1^H\mathbf{F}_1\right)}\\ &= \frac{P_2}{P_1M + N}\end{aligned} \quad (16)$$

where the second equality uses $\mathbb{E}[\boldsymbol{\Sigma}^2] = M\mathbf{I}$ because the diagonal entries of $\boldsymbol{\Sigma}^2$ is the unordered eigenvalues of a Wishart matrix $\mathbf{H}\mathbf{H}^H$ [28].

### B. Imperfect CSI Feedback via Vector Quantization

The structured precoding design requires channel knowledge at the transmitters. Generally, CSI can be obtained by each receiver through training. However, to obtain the channel information at the transmitters needs feedback. In our system, assuming perfect CSI at receivers (CSIR), the information of $\mathbf{V}$ needs to be sent back to the BS for determining $\mathbf{W}$ and the channel information $\mathbf{G}$ should be available at the RS through feedback for calculating $\mathbf{F}_1$. Notice that, since the knowledge of $\mathbf{H}$ is known at the RS, the information of $\mathbf{U}$ is always available at the RS for calculating $\mathbf{F}$.

In the limited feedback model shown in Fig. 1, the relay quantizes $\mathbf{V}$ to a finite number of bits (as an index) and then sends the index back to the BS. For the channels between the relay and users, each user quantizes its channel information $\mathbf{g}_k$ via vector quantization and feeds the quantization index back to the RS. We first consider the quantization of $\mathbf{V}$. For analytical simplicity in our study, we select the conventional vector quantization based method [19]–[21] for quantizing $\mathbf{V}$. Specifically, we use $B_1$ bits to quantize each column of $\mathbf{V}$ by a codebook $\mathcal{C}_1$ with $2^{B_1}$ unit-norm $M \times 1$ complex vector codewords. The quantization of $\mathbf{v}_k$ is obtained by performing

$$\hat{\mathbf{v}}_k = \arg \max_{\mathbf{c}_j \in \mathcal{C}_1,\, j=1,\cdots,2^{B_1}} \left| \mathbf{v}_k^H \mathbf{c}_j \right|^2. \quad (17)$$

With a quantized version of $\mathbf{V}$ as $\widehat{\mathbf{V}} = [\hat{\mathbf{v}}_1 \cdots \hat{\mathbf{v}}_N]$, a limited feedback version of the BS precoding matrix in (11) becomes

$$\mathbf{W} = \widehat{\mathbf{V}}. \quad (18)$$

For the downlink channel from the RS to users, each user $k$ quantizes its channel by using the same vector quantization method as that described in [20]

$$\hat{\mathbf{g}}_k = \arg \max_{\mathbf{c}_j \in \mathcal{C}_2,\, j=1,\cdots,2^{B_2}} \left| \mathbf{g}_k^H \mathbf{c}_j \right|^2 \quad (19)$$

where $B_2$ is the number of quantization bits and $\mathcal{C}_2$ is a codebook containing $2^{B_2}$ unit norm $N \times 1$ complex vector codewords. After obtaining $\widehat{\mathbf{G}} = [\hat{\mathbf{g}}_1 \cdots \hat{\mathbf{g}}_N]^H$ at the RS through feedback, the precoding matrix in (12) can be calculated as

$$\mathbf{F} = \widehat{\mathbf{F}} \mathbf{U}^H \quad (20)$$

where $\widehat{\mathbf{F}} = [\hat{\mathbf{f}}_1 \cdots \hat{\mathbf{f}}_N]$ is the ZFBF matrix calculated according to $\widehat{\mathbf{G}}$. Following (18) and (20), the joint BS and RS precoding scheme can now be implemented by utilizing the quantized CSI feedback, instead of the perfect channel knowledge.

## IV. THROUGHPUT ANALYSIS UNDER QUANTIZED CSI FEEDBACK

This section addresses the achievable throughput of the proposed limited feedback linear precoding scheme in a MIMO relaying BC. We first characterize the throughput loss as a function of the quantization quality of both two-hop links, i.e., $B_1$ and $B_2$, by comparing the limited feedback scheme with the approach with perfect CSI. Since the rate loss has shown to be an increasing function of system SNRs $P_1$ and $P_2$, we see that the system performance is interference-limited at high SNRs by deriving an upper bound of the achievable rate for infinite $P_1$ and $P_2$ with fixed CSI quantization quality.[1]

### A. An Upper Bound on the Rate Loss

Consider the linear precoding scheme with quantized feedback in (18) and (20), the received symbols in (7) can be rewritten by

$$\mathbf{y} = \sqrt{\rho_1 \rho_2} \mathbf{G} \widehat{\mathbf{F}} \mathbf{\Sigma} \mathbf{V}^H \widehat{\mathbf{V}} \mathbf{x}_s + \sqrt{\rho_2} \mathbf{G} \widehat{\mathbf{F}} \tilde{\mathbf{n}} + \mathbf{z}. \quad (21)$$

[1]Since both the transmit symbol and the noise are assumed to be of unit variance, we use the transmit power to denote SNR in this study.

Accordingly, the achieved SINR in (8) at user $k$ becomes

$$\gamma_k^Q = \frac{\left| \mathbf{g}_k^H \widehat{\mathbf{F}} \mathbf{\Sigma} \mathbf{V}^H \hat{\mathbf{v}}_k \right|^2}{\sum_{j \neq k} \left| \mathbf{g}_k^H \widehat{\mathbf{F}} \mathbf{\Sigma} \mathbf{V}^H \hat{\mathbf{v}}_j \right|^2 + \frac{1}{\rho_1} \left\| \mathbf{g}_k^H \widehat{\mathbf{F}} \right\|^2 + \frac{1}{\rho_1 \rho_2}}. \quad (22)$$

Then, from (14) and (22), the ergodic achievable rate with perfect CSI and quantized CSI feedback are

$$R_P = \frac{N}{2} \mathbb{E}\left[\log\left(1 + \gamma_k^P\right)\right] \quad (23)$$

and

$$R_Q = \frac{N}{2} \mathbb{E}\left[\log\left(1 + \gamma_k^Q\right)\right] \quad (24)$$

respectively.

*Theorem 1:* The achievable rate loss per user due to quantized feedback in the MIMO relay downlink can be upper bounded by

$$\begin{aligned} \Delta R &= \frac{1}{N}(R_P - R_Q) \\ &\leq \frac{1}{2} \log\left(1 + \mathbb{E}\left[\frac{\sqrt{\epsilon_k}}{\lambda_{\min}(\mathbf{Q})}\right]\right) \\ &\quad + \frac{1}{2} \log\left(1 + \rho_2(N-1)\left(\rho_1 2^{-\frac{B_1}{M-1}} + (1+\rho_1 M) 2^{-\frac{B_2}{N-1}}\right)\right) \end{aligned} \quad (25)$$

where $\epsilon_k$ is the quantization error of vector $\mathbf{v}_k$ defined in (49), $\lambda_{\min}(\cdot)$ is the minimum eigenvalue of a matrix, and matrix $\mathbf{Q}$ is defined by

$$\mathbf{Q} = \mathbf{V}^H \widehat{\mathbf{V}} \widehat{\mathbf{V}}^H \mathbf{V}$$

which only depends on the quantization level $B_1$ of the BS-to-RS channel.

*Proof:* See Appendix B. ∎

From Theorem 1 it follows that the rate loss is bounded from above by the sum of two components. The first sum component in (25) only depends on the CSI quantization for the BS-to-RS channel, while the second term is an increasing function of the system SNR $P_1$ and $P_2$. Since the value of the first term relies only on the quantization level $B_1$, we conclude that the second term in (25) dominates the rate loss at high SNR regime. Therefore, for high SNRs, the upper bound on the rate loss can be alternatively rewritten as

$$\begin{aligned} \Delta R &\leq \Delta R_{High} \\ &= \frac{1}{2} \log\Big(1 + \rho_2(N-1) \\ &\quad \times \left(\rho_1 2^{-\frac{B_1}{M-1}} + (1+\rho_1 M) 2^{-\frac{B_2}{N-1}}\right)\Big) + O(1). \end{aligned} \quad (26)$$

In summary, we have derived a simple expression of the rate loss upper bound which will be shown useful in the subsequent development. In order to verify the derived bound, Fig. 2 plots the derived upper bound $\Delta R_{\text{High}}$ for a system with $M = 4$ and $N = 2$ as a function of the number of feedback bits, along with the exact values of $\Delta R$ obtained by numerical methods. This figure verifies the accuracy of the upper bound versus the increasing values of both $B_1$ and $B_2$. The $O(1)$ term in (26) was neglected when calculating the theoretic bound $\Delta R_{\text{High}}$. From the results, it is observed that the upper bound gets tighter as the number of feedback bits increases.

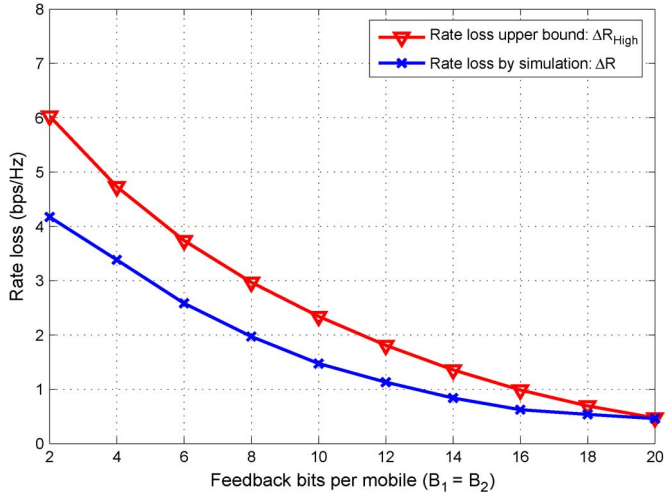

Fig. 2. Accuracy of the rate loss upper bound for $M = 4$ and $N = 2$ with $P_1 = P_2 = 25$ dB.

### B. Rate Upper Bound by Interference-Limited Effect

From the bound derived above, it is observed that the rate loss increases with the system SNRs $P_1$ and $P_2$. Therefore, it is also of interest to examine how the system performs at high SNRs with fixed feedback quality. The following result characterizes an upper bound for the achievable rate when the system transmit power constraints becomes unlimited.

*Theorem 2:* The achievable rate of the system using quantized feedback with either fixed $B_1$ or fixed $B_2$ is bounded from above as the system transmit power $P_1$ and $P_2$ becomes unlimited:

$$\frac{2}{N}R_Q(B_1) \leq \frac{2}{N}R_{U1}^{B_2 \to \infty}$$
$$= \log\left(1 - \frac{M-N}{M}2^{-\frac{B_1}{M-1}}\right) + \frac{\log e}{M-1}\sum_{k=1}^{2^{B_1}}\frac{1}{k}$$
$$+ \log e \sum_{k=1}^{M-2}\frac{1}{k} + \frac{\log e}{N-1} + c \quad (27)$$

and

$$\frac{2}{N}R_Q(B_2) \leq \frac{2}{N}R_{U2}^{B_1 \to \infty}$$
$$= \log\left(1 + (N-1)2^{-\frac{B_2}{N-1}}\right) + \frac{\log e}{N-1}\sum_{k=1}^{2^{B_2}}\frac{1}{k} + c \quad (28)$$

where $R_{U1}$ is the upper bound given as a function of $B_1$ with perfect relay-to-source feedback (infinite $B_2$), while $R_{U2}$ is an upper bound with respect to $B_2$ with perfect user-to-relay feedback (infinite $B_1$), and

$$c = \log\frac{M}{N(N-1)} + \log e \sum_{k=1}^{N-2}\frac{1}{k} - \frac{\log e}{N}\sum_{k=0}^{N-1}\psi(M-k)$$

is a constant independent of both $B_1$ and $B_2$, where $\psi(\cdot)$ is Euler's digamma function [30].

*Proof:* By symmetry, the achievable rate of the limited feedback system is bounded from above by

$$\frac{2}{N}R_Q$$
$$= \mathbb{E}\left[\log\left(1 + \frac{\left|\mathbf{g}_k^H\widehat{\mathbf{F}}\boldsymbol{\Sigma}\mathbf{V}^H\hat{\mathbf{v}}_k\right|^2}{\sum_{j\neq k}\left|\mathbf{g}_k^H\widehat{\mathbf{F}}\boldsymbol{\Sigma}\mathbf{V}^H\hat{\mathbf{v}}_j\right|^2 + \frac{1}{\rho_1}\left\|\mathbf{g}_k^H\widehat{\mathbf{F}}\right\|^2 + \frac{1}{\rho_1\rho_2}}\right)\right]$$

$$\leq \mathbb{E}\left[\log\left(1 + \frac{\left|\mathbf{g}_k^H\widehat{\mathbf{F}}\boldsymbol{\Sigma}\mathbf{V}^H\hat{\mathbf{v}}_k\right|^2}{\sum_{j\neq k}\left|\mathbf{g}_k^H\widehat{\mathbf{F}}\boldsymbol{\Sigma}\mathbf{V}^H\hat{\mathbf{v}}_j\right|^2}\right)\right] \quad (29)$$

$$= \mathbb{E}\left[\log\left(\left\|\tilde{\mathbf{g}}_k^H\widehat{\mathbf{F}}\boldsymbol{\Sigma}\mathbf{V}^H\widehat{\mathbf{V}}\right\|^2\right)\right]$$
$$- \mathbb{E}\left[\log\left(\sum_{j\neq k}\left|\tilde{\mathbf{g}}_k^H\widehat{\mathbf{F}}\boldsymbol{\Sigma}\mathbf{V}^H\hat{\mathbf{v}}_j\right|^2\right)\right] \quad (30)$$

$$\leq \log\left(\mathbb{E}\left[\left\|\tilde{\mathbf{g}}_k^H\widehat{\mathbf{F}}\boldsymbol{\Sigma}\mathbf{V}^H\widehat{\mathbf{V}}\right\|^2\right]\right)$$
$$- \mathbb{E}\left[\log\left(\frac{\sum_{j\neq k}\left|\tilde{\mathbf{g}}_k^H\widehat{\mathbf{F}}\boldsymbol{\Sigma}\mathbf{V}^H\hat{\mathbf{v}}_j\right|^2}{N-1}\right)\right] - \log(N-1) \quad (31)$$

$$\leq \log\left(\mathbb{E}\left[\left\|\tilde{\mathbf{g}}_k^H\widehat{\mathbf{F}}\boldsymbol{\Sigma}\mathbf{V}^H\widehat{\mathbf{V}}\right\|^2\right]\right)$$
$$- \frac{1}{N-1}\mathbb{E}\left[\sum_{j\neq k}\log\left(\left|\tilde{\mathbf{g}}_k^H\widehat{\mathbf{F}}\boldsymbol{\Sigma}\mathbf{V}^H\hat{\mathbf{v}}_j\right|^2\right)\right] - \log(N-1) \quad (32)$$

$$= \log\left(\mathbb{E}\left[\left\|\tilde{\mathbf{g}}_k^H\widehat{\mathbf{F}}\boldsymbol{\Sigma}\mathbf{V}^H\widehat{\mathbf{V}}\right\|^2\right]\right)$$
$$- \mathbb{E}\left[\log\left(\left|\tilde{\mathbf{g}}_k^H\widehat{\mathbf{F}}\boldsymbol{\Sigma}\mathbf{V}^H\hat{\mathbf{v}}_j\right|^2\right)\right] - \log(N-1) \quad (33)$$

where (29) omits two positive terms $(1/\rho_1)\|\mathbf{g}_k^H\widehat{\mathbf{F}}\|^2$ and $1/\rho_1\rho_2$ which are small for large $P_1$ and $P_2$, equality (30) uses the decomposition $\mathbf{g}_k = \|\mathbf{g}_k\|\tilde{\mathbf{g}}_k$ and eliminates common $\|\mathbf{g}_k\|^2$ in the two terms, inequalities (31) and (32) are followed by applying Jensen's inequality to the first term and the second term in the corresponding expressions, and the last expression (33) holds because of the fact that the expectation of the term $\log(|\tilde{\mathbf{g}}_k^H\widehat{\mathbf{F}}\boldsymbol{\Sigma}\mathbf{V}^H\hat{\mathbf{v}}_j|^2)$ is the same for different $j \neq k$. To obtain a closed-form expression of the rate upper bound, we need to use the results in Appendix C where the expectation terms in (33) are calculated separately.

We consider the case with fixed $B_1$ and perfect CSI feedback from users to relay. By substituting (79) and (83) (see Appendix C) into (33) and let $B_2 \to \infty$, the rate upper bound in (27) is obtained as a function of $B_1$. Consider now the case with fixed $B_2$ and perfect CSI feedback from relay to the BS. The upper bound in (28) is obtained by substituting (79) and (80) (see Appendix C) into (33) and let $B_1 \to \infty$, and the proof is complete. ∎

Theorem 2 implies that the system rate is bounded from above when $P_1$ and $P_2$ grow to infinity so long as one of the feedback quality $B_1$ and $B_2$ is a finite number. Fig. 3 plots the derived rate bounds as well as the numerical results of the achievable rate with $M = 4$ and $N = 2$ at high SNRs. Notice that we tested in this figure very high SNRs in order to verify the derived bounds although the high SNR values seem impractical in current wireless applications. These bounds show the performance limit of the relay system due to multiuser interference resulting from CSI quantization errors. More specifically, this phenomenon is called "interference-limited effect" in [20] due to the fact that both the signal power and the interference power in the limited feedback system increase linearly with the system transmit power. Moreover, by comparing the rate bounds $R_{U1}$ and $R_{U2}$, it is found that improving the feedback quality of one of the two-hop channels, i.e., increasing $B_1$ or $B_2$, has different impacts on enhancing the system throughput if the feedback quality of the other channel is assumed to be perfectly accurate. We shall discuss the problem of controlling feedback quality of both two-hop channels in the next section.

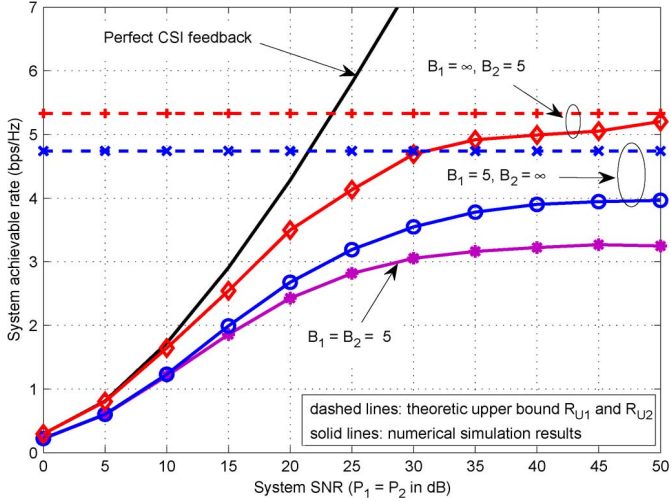

Fig. 3. Interference-limited effect in the MIMO Relaying BC using quantized feedback with $M = 4$ and $N = 2$ for high system SNRs.

## V. Feedback Quality Control for Two-Hop Channels

### A. Feedback Quality Control

We see from the results deduced above that if the quantization quality of either $B_1$ or $B_2$ is fixed, the rate loss becomes unbounded when the system SNR $P_1$ or $P_2$ increases to infinity. In this case, the system performance may eventually be overwhelmed by the multichannel (multiuser) interference due to the quantization error. This phenomenon is called "interference-limited effect" in the limited feedback MIMO BC without relay [20]. In order to mitigate this effect, one needs to increase the feedback quality with the system SNR. In what follows, we determine how the quantization quality including $B_1$ and $B_2$ must grow with $P_1$ and $P_2$ so that the rate loss becomes bounded from above by a given value. The next theorem offers a sufficient condition for the system to maintain a desired rate loss upper bound.

*Theorem 3:* In order to maintain the upper bounded rate loss no larger than $(1/2) \log b$ per user for high SNRs, it is sufficient to scale the numbers of CSI quantization bits per user, i.e., $B_1$ and $B_2$, according to

$$\frac{B_1}{M-1} = \log_2 P_2 - \log_2 \left( M + \frac{N}{P_1} \right) + \alpha \quad (34)$$

and

$$\frac{B_2}{N-1} = \log_2 P_2 + \alpha \quad (35)$$

where

$$\alpha = \log_2 \frac{2(N-1)}{N(b-1)}$$

is a value depends only on the number of remote users $N$ and parameter $b$ for a desired rate loss.

*Proof:* We consider the simple expression for the upper bound on the rate loss given by (26). In order to control the per user rate loss within a constant gap, we choose $B_1$ and $B_2$ by making the rate loss upper bound equal to $(1/2) \log b$. From (26), it follows that

$$\frac{1}{2} \log \left( 1 + \rho_2(N-1) \left( \rho_1 2^{-\frac{B_1}{M-1}} + (1+\rho_1 M) 2^{-\frac{B_2}{N-1}} \right) \right) + O(1) = \frac{1}{2} \log b. \quad (36)$$

Hence, by omitting the $O(1)$ term in (36) for high SNRs, a more tractable condition for $B_1$ and $B_2$ is given by

$$\frac{1}{2} \log \left( 1 + \rho_2(N-1) \left( \rho_1 2^{-\frac{B_1}{M-1}} + (1+\rho_1 M) 2^{-\frac{B_2}{N-1}} \right) \right) = \frac{1}{2} \log b \quad (37)$$

which implies that

$$2^{-\frac{B_1}{M-1}} + \left( M + \frac{1}{\rho_1} \right) 2^{-\frac{B_2}{N-1}} = \frac{b-1}{\rho_1 \rho_2 (N-1)}. \quad (38)$$

This is accurate and effective for an increasing SNR due to two reasons. First, the omitted term is a constant value under a fixed $B_1$ and is independent of the system SNR. To control the rate loss gap for a growing SNR, the left-hand side of (37), which increases with both $P_1$ and $P_2$, plays an essential role in controlling the entire rate loss. Secondly, if $B_1$ and $B_2$ are scaled according to the condition in (38) for increasing $P_1$ and $P_2$, the exact value of $O(1)$ also decreases to zero with increasing $B_1$, which guarantees the accuracy of this sufficient condition for high SNRs in our system.

Subsequently, in order to obtain a more meaningful expression for the feedback quality control, a new variable $\theta \in (0, 1)$ is introduced to help transform (38) to the following equivalent conditions:

$$2^{-\frac{B_1}{M-1}} = \frac{\theta(b-1)}{\rho_1 \rho_2 (N-1)}$$

$$\left( M + \frac{1}{\rho_1} \right) 2^{-\frac{B_2}{N-1}} = \frac{(1-\theta)(b-1)}{\rho_1 \rho_2 (N-1)} \quad (39)$$

where $\theta$ can be any value in (0, 1) and different values of $\theta$ may result in different solutions to $B_1$ and $B_2$. By solving the above equations in terms of $\theta$, $b$, $\rho_1$, and $\rho_2$ and then substituting (15) and (16) into the solutions, the conditions in (39) lead to the following concise expressions for controlling $B_1$ and $B_2$:

$$\frac{B_1}{M-1} = \log_2(\rho_1\rho_2) + \log_2(N-1) - \log_2(\theta(b-1))$$
$$= \log_2 P_2 - \log_2\left(M + \frac{N}{P_1}\right) + \log_2\frac{N-1}{\theta(b-1)N} \quad (40)$$

and

$$\frac{B_2}{N-1} = \log_2(\rho_1\rho_2) + \log_2\frac{N-1}{(1-\theta)(b-1)} + \log_2\left(M + \frac{1}{\rho_1}\right)$$
$$= \log_2 P_2 + \log_2\frac{N-1}{(1-\theta)(b-1)N}. \quad (41)$$

The above results imply that, for any given $b$, there can be different combinations of $B_1$ and $B_2$ to make condition (37) valid. In particular, choosing $\theta = 0.5$ in (40) and (41) leads to (34) and (35), respectively. ∎

From Theorem 3, several important and interesting conclusions on the issue of controlling the feedback quality can be made.

- In the limited feedback relay BC, in order to maintain the rate loss within a predetermined gap with respect to the perfect CSI case, *both $B_1$ and $B_2$ of the two-hop feedback channels should grow with the relay transmit power $P_2$ according to* (34) and (35), *respectively*. However, for an increasing BS transmit power $P_1$, only the number of relay-to-BS feedback bits $B_1$ needs to be increased.

- Although both quantization quality of two hop channels should increase with the transmit power, the power constraints $P_1$ and $P_2$ have different effects on scaling $B_1$ and $B_2$. Based on Theorem 3, it can be concluded that *the scaled $B_1$ is eventually upper bounded when $P_1$ tends to infinity and $P_2$ is fixed*. That is,

$$B_1|_{P_1\to\infty} = (M-1)\log_2 P_2$$
$$+ (M-1)\log_2\frac{2(N-1)}{(b-1)MN}. \quad (42)$$

- Consider a scenario where $B_1$ and $B_2$ are scaled with an increasing $P_2$ and fixed $P_1$. From Theorem 3, we have

$$B_1 \approx \frac{M-1}{3}P_2^{\text{dB}} \underbrace{-(M-1)\log_2\left(M+\frac{N}{P_1}\right) + \alpha}_{\text{constant offset with fixed }P_1} \quad (43)$$

and

$$B_2 \approx \frac{N-1}{3}P_2^{\text{dB}} + \alpha \quad (44)$$

where $P_2^{\text{dB}} = 10\log_{10} P_2$ is $P_2$ in decibels. This suggests that *for a fixed $P_1$, both quantization quality $B_1$ and $B_2$ of the two-hop channels should grow linearly with $P_2$ in decibels, which are unbounded*.

It is important to note that the above conditions in Theorem 3 is not necessary for all cases since it is obtained by using our derived upper bound to the rate loss. However, for high SNR cases, feedback quality control according to Theorem 3 is suggested to guarantee a desired rate loss level. Numerical results will be presented in Section V-C to demonstrate that the strategy for scaling $B_1$ and $B_2$ as stated in Theorem 3 behaves well in a variety of cases.

### B. Optimizing $\theta$ via Sum Feedback Minimization

In Theorem 3, the strategy of scaling $B_1$ and $B_2$ corresponds to $\theta = 0.5$. From (40) and (41), given the transmit power constraints with fixed $b$, different choices of $\theta$ may produce different combinations of $B_1$ and $B_2$, thus different sum feedback rates of both two-hop channels. Therefore, it is of interest to find an optimized $\theta$ such that the overall feedback rate of the system is minimized. In what follows, we show that $\theta = 0.5$ is indeed the optimal choice for the system with $M = N$.

Using (40) and (41), the sum feedback rate of the two-hop channels is calculated by

$$B_{\text{total}} = N(B_1 + B_2)$$
$$= N\left((M-1)\left(\log_2 P_2 - \log_2\left(M+\frac{N}{P_1}\right)\right.\right.$$
$$\left.+ \log_2\frac{N-1}{\theta(b-1)N}\right) + (N-1)$$
$$\left.\times\left(\log_2 P_2 + \log_2\frac{N-1}{(1-\theta)(b-1)N}\right)\right). \quad (45)$$

The problem of minimizing the overall feedback rate can be formulated as

$$\begin{array}{cl}\underset{\theta}{\text{minimize}} & B_{\text{total}} \\ \text{subject to}: & 0 < \theta < 1.\end{array} \quad (46)$$

Using the expression in (45) and after some manipulations, the problem in (85) is reduced to

$$\begin{array}{cl}\underset{\theta}{\text{maximize}} & \theta^{M-1}(1-\theta)^{N-1} \\ \text{subject to}: & 0 < \theta < 1.\end{array} \quad (47)$$

The above problem admits a closed-form solution as

$$\theta^* = \frac{M-1}{M+N-2} \quad (48)$$

and the solution is both unique and globally optimal, see Appendix D for a proof of this fact. It follows that the optimal solution $\theta^*$ only depends on the antenna configurations at the BS and relay. In particular, when $M = N$, $\theta^* = 0.5$ is the optimum value that minimizes the overall feedback rate.

### C. Numerical Results

We now present some numerical results for a system with $M = 4$ and $N = 2, 4$. Fig. 4 shows the achievable rate of the limited feedback system with fixed $B_1$ and $B_2$, comparing with the system using perfect CSI feedback. Although the achievable rate by quantized feedback is close to that of the system with perfect CSI in the low SNR region, it is quite pronounced that the performance becomes bounded from above in the high SNR region. Moreover, by comparing the results of different system configurations, it can be observed that high quality feedback becomes more essential for a system with more antennas (or serving more users) at high SNRs.

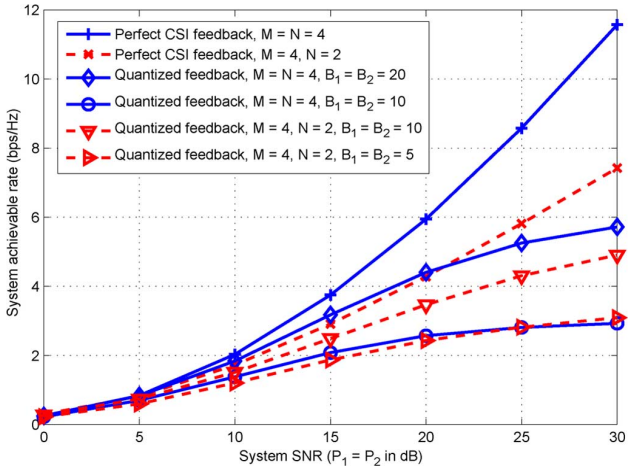

Fig. 4. Achievable sum rate of the MIMO relaying BC using quantized feedback.

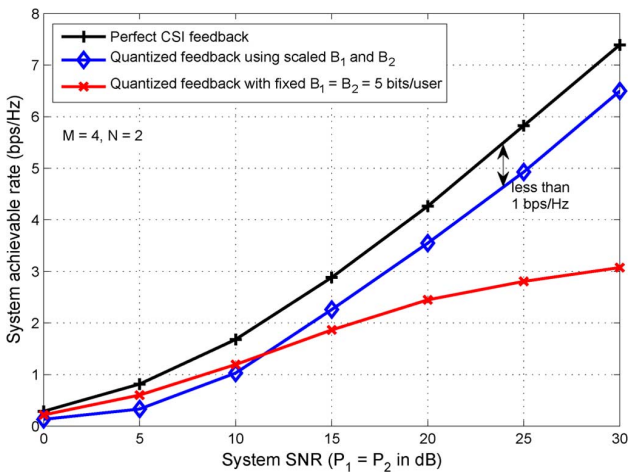

Fig. 5. MIMO relaying BC using quantized feedback with $M = 4$ and $N = 2$.

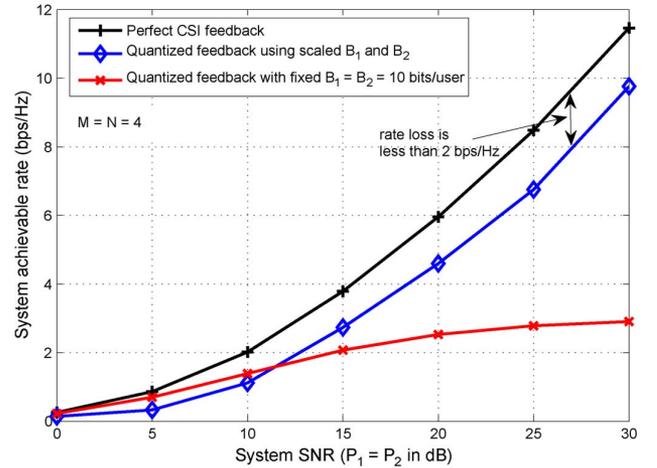

Fig. 6. MIMO relaying BC using quantized feedback with $M = 4$ and $N = 4$.

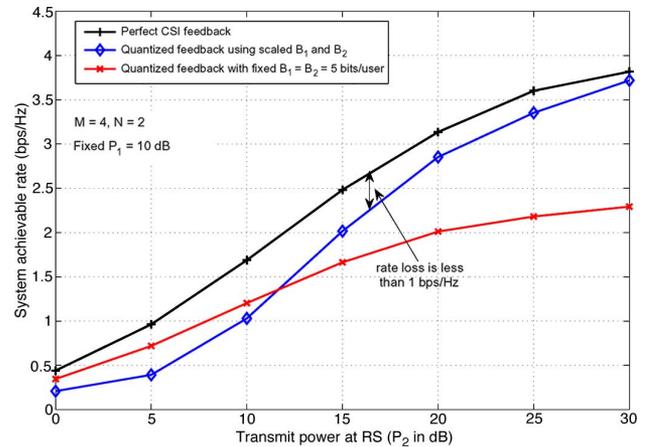

Fig. 7. MIMO relaying BC using quantized feedback with $M = 4$, $N = 2$, and fixed $P_1 = 10$ dB.

In order to let the achievable rate grow with the system SNR, we assumed the system using scaled $B_1$ and $B_2$ with increasing system SNRs $P_1$ and $P_2$. Specifically, the numbers of feedback bits $B_1$ and $B_2$ were taken from (34) and (35) in Theorem 3 so that the maximal rate loss due to quantized feedback is guaranteed. The value of $b$ was set to 2, resulting in a per user rate loss bounded from above by $(1/2)\log_2 b = 0.5$ b/s/Hz. Given $N$ users in the system, the overall system rate loss is accordingly bounded from above by $0.5N$ b/s/Hz.

In Figs. 5–8, the throughput achieved by using perfect CSI, quantized feedback using scaled $B_1$ and $B_2$, and quantized feedback with fixed $B_1$ and $B_2$ are shown for comparison. Fig. 5 depicts the results for a system with $M = 4$ and $N = 2$ as a function of increasing transmit power at both the BS source and relay, i.e., $P_1 = P_2$ ranging from 0 to 30 dB. As expected, it shows that the overall system rate loss is always controlled to remain within 1 b/s/Hz across the whole tested SNR regime. For a system with fixed $B_1$ and $B_2$, however, the rate loss due to quantized feedback is unbounded and grows with $P_1$ and $P_2$, hence resulting in an upper bounded system rate when $P_1$ and

$P_2$ increase. Fig. 6 demonstrates the results for $M = N = 4$. By using scaled $B_1$ and $B_2$, the system rate grows with the system SNR and the rate loss is always guaranteed to remain within 2 b/s/Hz with increasing transmit power $P_1$ and $P_2$. Notice that although the scaled $B_1$ and $B_2$ in Theorem 3 are derived by omitting the $O(1)$ term for large SNRs, the results in both figures verify its accuracy and effectiveness for different tested cases. Moreover, we evaluated a system with fixed $P_1$ while increasing $P_2$ from 0 to 30 dB and the results are shown in Fig. 7. Since the BS transmit power is fixed, the system rate is bounded from above even with perfect CSI feedback when $P_2$ becomes large. In this case, the rate loss by quantized feedback is well controlled by using scaled $B_1$ and $B_2$. As expected, however, when the feedback quality $B_1$ and $B_2$ are fixed, still the rate loss is found to grow with $P_2$. Contrary to Fig. 7, we also show in Fig. 8 the comparison results under fixed $P_2$ and increasing $P_1$ from 0 to 30 dB. Since $P_2$ is fixed, the overall system performance is upper bounded when $P_1$ increases to large. However, by utilizing the proposed scaling $B_1$ and $B_2$ in our paper, the rate loss is well controlled within 1 b/s/Hz.

## VI. CONCLUSION

This paper studies a limited feedback linear precoding scheme in a MIMO relay-assisted downlink system. With a quantized CSI feedback employed, the performance degradation relative to the perfect CSI case is quantified by deriving an upper bound to the rate loss. As the transmit power grows large, the achievable rate has been shown to be eventually bounded from above due to the "interference-limited effect" in limited feedback multiuser systems. Therefore, in order to scale the achievable rate with system SNRs, a feedback quality control strategy has been developed. It is shown that both the quantization quality in the two-hop channels need to be increased with the transmit power at the relay. However, for growing transmit power at the BS, only the feedback quality from the relay-to-BS link should be increased. The problem of minimizing the sum feedback rate of both two-hop channels for the proposed feedback strategy is also addressed.

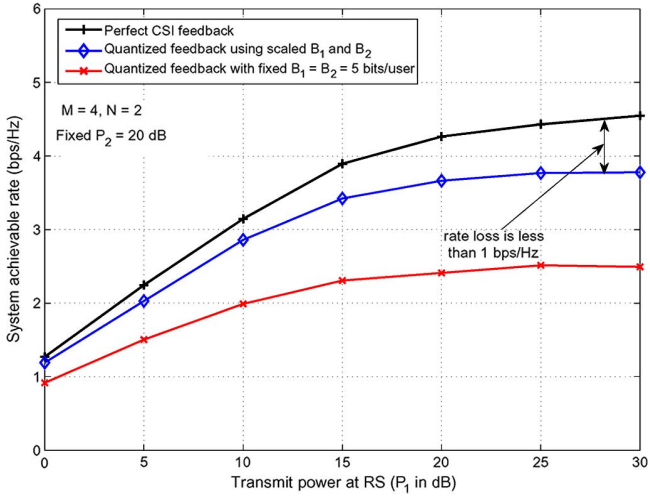

Fig. 8. MIMO relaying BC using quantized feedback with $M = 4$, $N = 2$, and fixed $P_2 = 20$ dB.

## APPENDIX A
## SOME PRELIMINARIES ON VECTOR QUANTIZATION PROPERTIES

In this appendix, we provide some preliminary results about the characteristics of quantization vectors which are shown useful in our paper. For brevity, define the quantization error of vector $\mathbf{v}_j$ by

$$\epsilon_j = 1 - \left|\mathbf{v}_j^H \hat{\mathbf{v}}_j\right|^2 \quad (49)$$

and let $\bar{\epsilon}$ denote the expected quantization error as follows:

$$\bar{\epsilon} = \mathbb{E}\left[1 - \left|\mathbf{v}_j^H \hat{\mathbf{v}}_j\right|^2\right] \quad (50)$$

which is the same for any specific vector $\mathbf{v}_j$ because they are identically distributed unit vectors. Moreover, since it is in general difficult to consider the optimal design of a codebook for some given metric [25], random vector quantization codebooks [20]–[23] are considered here.

*Lemma 1:* The random variable $\left|\mathbf{v}_i^H \hat{\mathbf{v}}_j\right|^2$ for any $i \neq j$ is statistically equivalent to the new variable which is the product of the vector quantization error $\epsilon_j$ and a $\text{beta}(1, M-2)$ random variable.

*Proof:* This result directly follows from [20, Lemma 2] because $\hat{\mathbf{v}}_j$ is the quantization of $\mathbf{v}_j$ which is orthogonal to $\mathbf{v}_i$. ∎

*Lemma 2:* For any quantization vector $\hat{\mathbf{v}}_k$, we have

$$\mathbb{E}_\mathbf{V}\left[\mathbf{V}^H \hat{\mathbf{v}}_k \hat{\mathbf{v}}_k^H \mathbf{V}\right]$$
$$= \text{diag}\left\{\frac{1}{M-1}\bar{\epsilon}, \cdots, \underbrace{(1-\bar{\epsilon})}_{k\text{th element}}, \frac{1}{M-1}\bar{\epsilon}, \cdots\right\} \quad (51)$$

where $\text{diag}\{\cdots\}$ is a diagonal matrix with given elements in the diagonal.

*Proof:* We prove the expectation of the matrix by separately considering the expectation of each $(i, j)$th element of the matrix as follows:

$$\mathbb{E}\left[\left[\mathbf{V}^H \hat{\mathbf{v}}_k \hat{\mathbf{v}}_k^H \mathbf{V}\right]_{i,j}\right] = \mathbb{E}\left[\mathbf{v}_i^H \hat{\mathbf{v}}_k \hat{\mathbf{v}}_k^H \mathbf{v}_j\right]. \quad (52)$$

- For $i = j = k$: The definition of $\bar{\epsilon}$ directly gives

$$\mathbb{E}\left[\left|\mathbf{v}_k^H \hat{\mathbf{v}}_k\right|^2\right] = 1 - \bar{\epsilon}. \quad (53)$$

- For $i = j \neq k$: By using Lemma 1, it gives

$$\mathbb{E}\left[\left|\mathbf{v}_i^H \hat{\mathbf{v}}_k\right|^2\right] = \mathbb{E}[\epsilon_k]\mathbb{E}[\text{beta}(1, M-2)] = \frac{\bar{\epsilon}}{M-1}. \quad (54)$$

- For $i \neq j$: In this case, at least one of the vector indexes $i$ and $j$ should be different from $k$. Hence, without loss of generality, consider the case with $j \neq k$. Since vector $\mathbf{v}_j$ is the $j$th column of a Haar matrix $[\mathbf{V}\ \mathbf{V}_r]$ which is uniformly distributed on the set of all $M \times M$ unitary matrices [28], vector $\mathbf{v}_j$ is isotropically distributed in the unit-norm $M \times 1$ vector space. Thus, the new vector of $\tilde{\mathbf{v}}_j = e^{-j\pi}\mathbf{v}_j$, which is still orthogonal with other vectors $\mathbf{v}_i$ ($i \neq j$) in the unitary matrix $\mathbf{V}$, follows the same distribution of $\mathbf{v}_j$. Then, we obtain

$$-\mathbb{E}\left[\mathbf{v}_i^H \hat{\mathbf{v}}_k \hat{\mathbf{v}}_k^H \mathbf{v}_j\right] = \mathbb{E}\left[e^{-j\pi}\mathbf{v}_i^H \hat{\mathbf{v}}_k \hat{\mathbf{v}}_k^H \mathbf{v}_j\right]$$
$$= \mathbb{E}\left[\mathbf{v}_i^H \hat{\mathbf{v}}_k \hat{\mathbf{v}}_k^H \tilde{\mathbf{v}}_j\right]$$
$$= \mathbb{E}\left[\mathbf{v}_i^H \hat{\mathbf{v}}_k \hat{\mathbf{v}}_k^H \mathbf{v}_j\right]$$

which implies

$$\mathbb{E}\left[\mathbf{v}_i^H \hat{\mathbf{v}}_k \hat{\mathbf{v}}_k^H \mathbf{v}_j\right] = 0. \quad (55)$$

Then, the Lemma follows by combing the results in (53)–(55). ∎

Notice that from Lemma 2, we can further obtain the following two useful results:

$$\sum_{j=1, j \neq k}^{N} \mathbb{E}\left[\mathbf{V}^H \hat{\mathbf{v}}_j \hat{\mathbf{v}}_j^H \mathbf{V}\right] = \text{diag}\left\{1 - \frac{M-N+1}{M-1}\bar{\epsilon}, \cdots, \right.$$
$$\left.\underbrace{\frac{N-1}{M-1}\bar{\epsilon}}_{k\text{th element}}, 1 - \frac{M-N+1}{M-1}\bar{\epsilon}, \cdots\right\} \quad (56)$$

and

$$\mathbb{E}[\mathbf{V}^H \hat{\mathbf{V}} \hat{\mathbf{V}}^H \mathbf{V}] = \sum_{j=1}^{N} \mathbb{E}_\mathbf{V}\left[\mathbf{V}^H \hat{\mathbf{v}}_j \hat{\mathbf{v}}_j^H \mathbf{V}\right]$$
$$= \left(1 - \frac{M-N}{M-1}\bar{\epsilon}\right)\mathbf{I}_N. \quad (57)$$

*Lemma 3:* For a matrix $\mathbf{A} = \sqrt{\epsilon_k}\mathbf{I} - (\mathbf{v}_k\mathbf{v}_k^H - \hat{\mathbf{v}}_k\hat{\mathbf{v}}_k^H)$ where $\mathbf{v}_k$ and $\hat{\mathbf{v}}_k$ are arbitrary unit-norm vectors, $\mathbf{A}$ is positive semidefinite, i.e., $\mathbf{A} \succeq \mathbf{0}$.

*Proof:* It is obvious that matrix $\mathbf{A}$ is Hermitian. From [29, Theorem 7.2.1], a Hermitian matrix is positive semidefinite if and only if all its eigenvalues of are nonnegative. The following will show that this condition always holds for $\mathbf{A}$.

Given $\hat{\mathbf{v}}_k$ is the quantization of the unit-norm vector $\mathbf{v}_k$, we begin with the decomposition [20]:

$$\hat{\mathbf{v}}_k = \sqrt{1-\epsilon_k}\mathbf{v}_k + \sqrt{\epsilon_k}\mathbf{s}_k \quad (58)$$

where $\mathbf{s}_k$ is a unit-norm vector in the null space of $\mathbf{v}_k$ and it is independent with the quantization error $\epsilon_k$. Define

$$\mathbf{B} = \mathbf{v}_k\mathbf{v}_k^H - \hat{\mathbf{v}}_k\hat{\mathbf{v}}_k^H. \quad (59)$$

By substituting (58), it yields

$$\begin{aligned}\mathbf{B} &= \mathbf{v}_k\mathbf{v}_k^H - (\sqrt{1-\epsilon_k}\mathbf{v}_k + \sqrt{\epsilon_k}\mathbf{s}_k)(\sqrt{1-\epsilon_k}\mathbf{v}_k + \sqrt{\epsilon_k}\mathbf{s}_k)^H \\ &= \epsilon_k\mathbf{v}_k\mathbf{v}_k^H - \sqrt{\epsilon_k(1-\epsilon_k)}\left(\mathbf{s}_k\mathbf{v}_k^H + \mathbf{v}_k\mathbf{s}_k^H\right) - \epsilon_k\mathbf{s}_k\mathbf{s}_k^H.\end{aligned} \quad (60)$$

If $\lambda$ is an eigenvalue of $\mathbf{B}$ with its corresponding eigenvector $\mathbf{x}$, then $\mathbf{B}\mathbf{x} = \lambda\mathbf{x}$. Thus, from (60), we have

$$\mathbf{v}_k^H\mathbf{B}\mathbf{x} = \epsilon_k\mathbf{v}_k^H\mathbf{x} - \sqrt{\epsilon_k(1-\epsilon_k)}\mathbf{s}_k^H\mathbf{x} = \lambda\mathbf{v}_k^H\mathbf{x} \quad (61)$$

where the property that $\mathbf{v}_k$ is perpendicular to $\mathbf{s}_k$, i.e., $\mathbf{v}_k^H\mathbf{s}_k = 0$, and $\mathbf{v}_k^H\mathbf{v}_k = 1$ are used. Similarly by multiplying $\mathbf{s}_k^H$ at both sides of $\mathbf{B}\mathbf{x} = \lambda\mathbf{x}$, we obtain

$$\mathbf{s}_k^H\mathbf{B}\mathbf{x} = -\sqrt{\epsilon_k(1-\epsilon_k)}\mathbf{v}_k^H\mathbf{x} - \epsilon_k\mathbf{s}_k^H\mathbf{x} = \lambda\mathbf{s}_k^H\mathbf{x}. \quad (62)$$

After some basic manipulations with (61) and (62), it yields the nonzero eigenvalues of $\mathbf{B}$ as

$$\lambda = \pm\sqrt{\epsilon_k}. \quad (63)$$

Thus far, all eigenvalues of $\mathbf{A}$ are proven to be nonnegative values because $\mathbf{A} = \sqrt{\epsilon_k}\mathbf{I} - \mathbf{B}$. ∎

## APPENDIX B
## PROOF OF THEOREM 1

From (14) and (22)–(24), the rate loss $\Delta R = R_P - R_Q$ can be written by (64), shown at the bottom of the page. Here, the last equality follows by the fact that $\mathbf{f}_k$ and $\hat{\mathbf{f}}_k$ follow the same distribution and both are independent of $\mathbf{g}_k$ [20]. For brevity, define $\Delta_1$ and $\Delta_2$ as the first and the second terms in the summation of (64), respectively. Then, by separately deriving the upper bounds to the two variables $\Delta_1$ and $\Delta_2$, we can obtain the desired bound on $\Delta R$.

Concerning $\Delta_1$, it can be upper bounded by (65)–(67), shown at the bottom of the next page, where (65) holds by eliminating two nonpositive terms in the numerator, inequality (66) uses $\sqrt{\epsilon_k}\mathbf{I} \succeq (\mathbf{v}_k\mathbf{v}_k^H - \hat{\mathbf{v}}_k\hat{\mathbf{v}}_k^H)$ given in Lemma 3, and the last inequality (67) follows by the generalized Rayleigh quotient theorem.

Subsequently, by applying Jensen's inequality, $\Delta_2$ can be upper bounded by

$$\begin{aligned}\Delta_2 &= \mathbb{E}\left[\log\left(1 + \frac{\frac{1}{\rho_1}\sum_{j\neq k}\left|\mathbf{g}_k^H\hat{\mathbf{f}}_j\right|^2 + \sum_{j\neq k}\left|\mathbf{g}_k^H\hat{\mathbf{F}}\boldsymbol{\Sigma}\mathbf{V}^H\hat{\mathbf{v}}_j\right|^2}{\frac{1}{\rho_1}\left|\mathbf{g}_k^H\hat{\mathbf{f}}_k\right|^2 + \frac{1}{\rho_1\rho_2}}\right)\right] \\ &\leq \mathbb{E}\left[\log\left(1 + \frac{\frac{1}{\rho_1}\sum_{j\neq k}\left|\mathbf{g}_k^H\hat{\mathbf{f}}_j\right|^2 + \sum_{j\neq k}\left|\mathbf{g}_k^H\hat{\mathbf{F}}\boldsymbol{\Sigma}\mathbf{V}^H\hat{\mathbf{v}}_j\right|^2}{\frac{1}{\rho_1\rho_2}}\right)\right] \\ &= \mathbb{E}\left[\log\left(1 + \rho_2\sum_{j\neq k}\left|\mathbf{g}_k^H\hat{\mathbf{f}}_j\right|^2 + \rho_1\rho_2\sum_{j\neq k}\left|\mathbf{g}_k^H\hat{\mathbf{F}}\boldsymbol{\Sigma}\mathbf{V}^H\hat{\mathbf{v}}_j\right|^2\right)\right] \\ &\leq \log\left(1 + \rho_2\sum_{j\neq k}\mathbb{E}\left[\left|\mathbf{g}_k^H\hat{\mathbf{f}}_j\right|^2\right] \right.\\ &\quad\left. + \rho_1\rho_2\sum_{j\neq k}\mathbb{E}\left[\left|\mathbf{g}_k^H\hat{\mathbf{F}}\boldsymbol{\Sigma}\mathbf{V}^H\hat{\mathbf{v}}_j\right|^2\right]\right).\end{aligned} \quad (68)$$

$$\begin{aligned}2\Delta R &= \mathbb{E}\left[\log\left(1 + \frac{\sigma_k^2\left|\mathbf{g}_k^H\mathbf{f}_k\right|^2}{\frac{1}{\rho_1}\left|\mathbf{g}_k^H\mathbf{f}_k\right|^2 + \frac{1}{\rho_1\rho_2}}\right) - \log\left(1 + \frac{\left|\mathbf{g}_k^H\hat{\mathbf{F}}\boldsymbol{\Sigma}\mathbf{V}^H\hat{\mathbf{v}}_k\right|^2}{\sum_{j\neq k}\left|\mathbf{g}_k^H\hat{\mathbf{F}}\boldsymbol{\Sigma}\mathbf{V}^H\hat{\mathbf{v}}_j\right|^2 + \frac{1}{\rho_1}\left\|\mathbf{g}_k^H\hat{\mathbf{F}}\right\|^2 + \frac{1}{\rho_1\rho_2}}\right)\right] \\ &= \mathbb{E}\left[\log\left(\frac{1}{\rho_1}\left|\mathbf{g}_k^H\mathbf{f}_k\right|^2 + \frac{1}{\rho_1\rho_2} + \sigma_k^2\left|\mathbf{g}_k^H\mathbf{f}_k\right|^2\right)\right] - \mathbb{E}\left[\log\left(\frac{1}{\rho_1}\left\|\mathbf{g}_k^H\hat{\mathbf{F}}\right\|^2 + \frac{1}{\rho_1\rho_2} + \left\|\mathbf{g}_k^H\hat{\mathbf{F}}\boldsymbol{\Sigma}\mathbf{V}^H\hat{\mathbf{V}}\right\|^2\right)\right] \\ &\quad + \mathbb{E}\left[\log\left(\frac{1}{\rho_1}\left\|\mathbf{g}_k^H\hat{\mathbf{F}}\right\|^2 + \frac{1}{\rho_1\rho_2} + \sum_{j\neq k}\left|\mathbf{g}_k^H\hat{\mathbf{F}}\boldsymbol{\Sigma}\mathbf{V}^H\hat{\mathbf{v}}_j\right|^2\right)\right] - \mathbb{E}\left[\log\left(\frac{1}{\rho_1}\left|\mathbf{g}_k^H\mathbf{f}_k\right|^2 + \frac{1}{\rho_1\rho_2}\right)\right] \\ &= \mathbb{E}\left[\log\left(1 + \frac{\sigma_k^2\left|\mathbf{g}_k^H\hat{\mathbf{f}}_k\right|^2 - \left\|\mathbf{g}_k^H\hat{\mathbf{F}}\boldsymbol{\Sigma}\mathbf{V}^H\hat{\mathbf{V}}\right\|^2 - \frac{1}{\rho_1}\sum_{j\neq k}\left|\mathbf{g}_k^H\hat{\mathbf{f}}_j\right|^2}{\frac{1}{\rho_1}\left\|\mathbf{g}_k^H\hat{\mathbf{F}}\right\|^2 + \frac{1}{\rho_1\rho_2} + \left\|\mathbf{g}_k^H\hat{\mathbf{F}}\boldsymbol{\Sigma}\mathbf{V}^H\hat{\mathbf{V}}\right\|^2}\right)\right] \\ &\quad + \mathbb{E}\left[\log\left(1 + \frac{\frac{1}{\rho_1}\sum_{j\neq k}\left|\mathbf{g}_k^H\hat{\mathbf{f}}_j\right|^2 + \sum_{j\neq k}\left|\mathbf{g}_k^H\hat{\mathbf{F}}\boldsymbol{\Sigma}\mathbf{V}^H\hat{\mathbf{v}}_j\right|^2}{\frac{1}{\rho_1}\left|\mathbf{g}_k^H\hat{\mathbf{f}}_k\right|^2 + \frac{1}{\rho_1\rho_2}}\right)\right].\end{aligned} \quad (64)$$

Since the two matrices $\mathbf{\Sigma}$ and $\mathbf{V}$ from SVD of $\mathbf{H}$ are independent and the two hop channels $\mathbf{H}$ and $\mathbf{G}$ are also independent with each other, we have

$$\sum_{j \neq k} \mathbb{E}\left[\left|\mathbf{g}_k^H \widehat{\mathbf{F}} \mathbf{\Sigma} \mathbf{V}^H \hat{\mathbf{v}}_j\right|^2\right]$$

$$= \mathbb{E}_{\mathbf{G},\mathbf{\Sigma}}\left[\mathbf{g}_k^H \widehat{\mathbf{F}} \mathbf{\Sigma} \left(\sum_{j \neq k} \mathbb{E}_{\mathbf{V}}\left[\mathbf{V}^H \hat{\mathbf{v}}_j \hat{\mathbf{v}}_j^H \mathbf{V}\right]\right) \mathbf{\Sigma} \widehat{\mathbf{F}}^H \mathbf{g}_k\right]$$

$$= \mathbb{E}_{\mathbf{G}}\left[\mathbf{g}_k^H \widehat{\mathbf{F}} \mathbb{E}_{\mathbf{\Sigma}}[\mathbf{\Sigma}^2] \left(\frac{(N-1)\bar{\epsilon}}{M-1} \mathbf{e}_k \mathbf{e}_k^H \right.\right.$$
$$\left.\left.+ \sum_{j \neq k}\left(1 - \frac{M-N+1}{M-1}\bar{\epsilon}\right) \mathbf{e}_j \mathbf{e}_j^H\right) \widehat{\mathbf{F}}^H \mathbf{g}_k\right] \quad (69)$$

$$= \frac{M(N-1)\bar{\epsilon}}{M-1} \mathbb{E}\left[\left|\mathbf{g}_k^H \hat{\mathbf{f}}_k\right|^2\right] + M\left(1 - \frac{M-N+1}{M-1}\bar{\epsilon}\right)$$
$$\times \sum_{j \neq k} \mathbb{E}\left[\left|\mathbf{g}_k^H \hat{\mathbf{f}}_j\right|^2\right] \quad (70)$$

$$= \frac{M(N-1)\bar{\epsilon}}{M-1} + MN\left(1 - \frac{M-N+1}{M-1}\bar{\epsilon}\right)\bar{\tau} \quad (71)$$

where, in (69) we use the result given in (56) (Appendix A) and $\mathbf{e}_i$ is the vector with 1 as the $i$th element and zeros elsewhere. Equality (70) uses $\mathbb{E}[\mathbf{\Sigma}^2] = M\mathbf{I}_N$ [28]. The final equality (71) uses the following expectation results [22]

$$\mathbb{E}_{\mathbf{G}}\left[\left|\mathbf{g}_k^H \hat{\mathbf{f}}_k\right|^2\right] = \mathbb{E}\left[\chi_{2N}^2\right] \mathbb{E}\left[\text{beta}(1, N-1)\right] = 1 \quad (72)$$

and

$$\mathbb{E}_{\mathbf{G}}\left[\left|\mathbf{g}_k^H \hat{\mathbf{f}}_j\right|^2\right] = \mathbb{E}\left[\chi_{2N}^2\right] \mathbb{E}\left[\text{beta}(1, N-2)\right] \mathbb{E}\left[1 - \left|\tilde{\mathbf{g}}_k^H \hat{\mathbf{g}}_k\right|^2\right]$$

$$= \frac{N}{N-1}\bar{\tau} \quad (73)$$

where $\tilde{\mathbf{g}}_k = \mathbf{g}_k/\|\mathbf{g}_k\|$ is the normalized channel vector and $\bar{\tau}$ is defined by

$$\bar{\tau} = \mathbb{E}\left[1 - \left|\tilde{\mathbf{g}}_k^H \hat{\mathbf{g}}_k\right|^2\right].$$

At this point, by substituting (71) and (73) into (68), the upper bound to $\Delta_2$ is rewritten by

$$\Delta_2 \leq \log\left(1 + \rho_2 N \bar{\tau} + \rho_1 \rho_2 \frac{M(N-1)\bar{\epsilon}}{M-1}\right.$$
$$\left.+ \rho_1 \rho_2 MN\left(1 - \frac{M-N+1}{M-1}\bar{\epsilon}\right)\bar{\tau}\right)$$

$$= \log\left(1 + \rho_1\rho_2 \frac{M(N-1)}{M-1}\bar{\epsilon} + \rho_2 N(1 + \rho_1 M)\bar{\tau}\right.$$
$$\left.- \rho_1\rho_2 \frac{MN(M-N+1)}{M-1}\bar{\epsilon}\bar{\tau}\right)$$

$$\leq \log\left(1 + \rho_1\rho_2(N-1)2^{-\frac{B_1}{M-1}}\right.$$
$$\left.+ \rho_2(N-1)(1+\rho_1 M)2^{-\frac{B_2}{N-1}}\right). \quad (74)$$

The last step is obtained by first omitting the negative second-order term of the quantization errors and then replacing the expected quantization errors by [22][2]:

$$\bar{\epsilon} = \frac{M-1}{M} 2^{-\frac{B_1}{M-1}}, \quad \bar{\tau} = \frac{N-1}{N} 2^{-\frac{B_2}{N-1}}. \quad (75)$$

[2]In fact, a more complex expression for the exact expected quantization error has been given in [21] for random vector quantization. However, to facilitate the performance analysis in our study, we here use a simpler result which has been shown in [22] to be a very accurate approximation for performance analysis.

---

$$\Delta_1 = \mathbb{E}\left[\log\left(1 + \frac{\sigma_k^2\left|\mathbf{g}_k^H \hat{\mathbf{f}}_k\right|^2 - \left|\mathbf{g}_k^H \widehat{\mathbf{F}}\mathbf{\Sigma}\mathbf{V}^H \hat{\mathbf{v}}_k\right|^2 - \sum_{j \neq k}\left|\mathbf{g}_k^H \widehat{\mathbf{F}}\mathbf{\Sigma}\mathbf{V}^H \hat{\mathbf{v}}_j\right|^2 - \frac{1}{\rho_1}\sum_{j \neq k}\left|\mathbf{g}_k^H \hat{\mathbf{f}}_j\right|^2}{\frac{1}{\rho_1}\left\|\mathbf{g}_k^H \widehat{\mathbf{F}}\right\|^2 + \frac{1}{\rho_1\rho_2} + \left\|\mathbf{g}_k^H \widehat{\mathbf{F}}\mathbf{\Sigma}\mathbf{V}^H \widehat{\mathbf{V}}\right\|^2}\right)\right]$$

$$\leq \mathbb{E}\left[\log\left(1 + \frac{\sigma_k^2\left|\mathbf{g}_k^H \hat{\mathbf{f}}_k\right|^2 - \left|\mathbf{g}_k^H \widehat{\mathbf{F}}\mathbf{\Sigma}\mathbf{V}^H \hat{\mathbf{v}}_k\right|^2}{\frac{1}{\rho_1}\left\|\mathbf{g}_k^H \widehat{\mathbf{F}}\right\|^2 + \frac{1}{\rho_1\rho_2} + \left\|\mathbf{g}_k^H \widehat{\mathbf{F}}\mathbf{\Sigma}\mathbf{V}^H \widehat{\mathbf{V}}\right\|^2}\right)\right] \quad (65)$$

$$= \mathbb{E}\left[\log\left(1 + \frac{\mathbf{g}_k^H \widehat{\mathbf{F}}\mathbf{\Sigma}\mathbf{V}^H \left(\mathbf{v}_k \mathbf{v}_k^H - \hat{\mathbf{v}}_k \hat{\mathbf{v}}_k^H\right)\mathbf{V}\mathbf{\Sigma}\widehat{\mathbf{F}}^H \mathbf{g}_k}{\frac{1}{\rho_1}\left\|\mathbf{g}_k^H \widehat{\mathbf{F}}\right\|^2 + \frac{1}{\rho_1\rho_2} + \left\|\mathbf{g}_k^H \widehat{\mathbf{F}}\mathbf{\Sigma}\mathbf{V}^H \widehat{\mathbf{V}}\right\|^2}\right)\right]$$

$$\leq \mathbb{E}\left[\log\left(1 + \sqrt{\epsilon_k}\frac{\mathbf{g}_k^H \widehat{\mathbf{F}}\mathbf{\Sigma}\mathbf{V}^H \mathbf{V}\mathbf{\Sigma}\widehat{\mathbf{F}}^H \mathbf{g}_k}{\left\|\mathbf{g}_k^H \widehat{\mathbf{F}}\mathbf{\Sigma}\mathbf{V}^H \widehat{\mathbf{V}}\right\|^2}\right)\right] \quad (66)$$

$$= \mathbb{E}\left[\log\left(1 + \sqrt{\epsilon_k}\frac{(\mathbf{\Sigma}\widehat{\mathbf{F}}^H \mathbf{g}_k)^H (\mathbf{\Sigma}\widehat{\mathbf{F}}^H \mathbf{g}_k)}{(\mathbf{\Sigma}\widehat{\mathbf{F}}^H \mathbf{g}_k)^H \mathbf{V}^H \widehat{\mathbf{V}}\widehat{\mathbf{V}}^H \mathbf{V}(\mathbf{\Sigma}\widehat{\mathbf{F}}^H \mathbf{g}_k)}\right)\right]$$

$$\leq \mathbb{E}\left[\log\left(1 + \frac{\sqrt{\epsilon_k}}{\lambda_{\min}(\mathbf{V}^H \widehat{\mathbf{V}}\widehat{\mathbf{V}}^H \mathbf{V})}\right)\right] \quad (67)$$

Thus far, by applying Jensen's inequality to (67) and from (74), we finally obtain the upper bound to the rate loss $\Delta R$ as

$$2\Delta R = \Delta_1 + \Delta_2$$
$$\leq \log\left(1 + \mathbb{E}\left[\frac{\sqrt{\epsilon_k}}{\lambda_{\min}(\mathbf{V}^H\widehat{\mathbf{V}}\widehat{\mathbf{V}}^H\mathbf{V})}\right]\right)$$
$$+ \log\left(1 + \rho_2(N-1)\left(\rho_1 2^{-\frac{B_1}{M-1}} + (1+\rho_1 M)2^{-\frac{B_2}{N-1}}\right)\right) \quad (76)$$

which completes the proof.

## APPENDIX C
## CALCULATIONS FOR PROOF OF THEOREM 2

This appendix provides some expectation results for the proof of Theorem 2. First, the following expectation is given by

$$\mathbb{E}\left[\left\|\tilde{\mathbf{g}}_k^H\widehat{\mathbf{F}}\boldsymbol{\Sigma}\mathbf{V}^H\widehat{\mathbf{V}}\right\|^2\right]$$
$$= \mathbb{E}_{\mathbf{G}}\left[\tilde{\mathbf{g}}_k^H\widehat{\mathbf{F}}\mathbb{E}_{\boldsymbol{\Sigma}}\left[\boldsymbol{\Sigma}\mathbb{E}_{\mathbf{V}}[\mathbf{V}^H\widehat{\mathbf{V}}\widehat{\mathbf{V}}^H\mathbf{V}]\boldsymbol{\Sigma}\right]\widehat{\mathbf{F}}^H\tilde{\mathbf{g}}_k\right]$$
$$= M\left(1 - \frac{M-N}{M-1}\bar{\epsilon}\right)\mathbb{E}_{\mathbf{G}}\left[\tilde{\mathbf{g}}_k^H\widehat{\mathbf{F}}\widehat{\mathbf{F}}^H\tilde{\mathbf{g}}_k\right] \quad (77)$$
$$= M\left(1 - \frac{M-N}{M-1}\bar{\epsilon}\right)\sum_{j=1}^N \mathbb{E}_{\mathbf{G}}\left[\left|\tilde{\mathbf{g}}_k^H\hat{\mathbf{f}}_j\right|^2\right]$$
$$= M\left(1 - \frac{M-N}{M-1}\bar{\epsilon}\right)\left(\frac{1}{N} + \bar{\tau}\right) \quad (78)$$

where (77) uses (57) in Appendix A and $\mathbb{E}[\boldsymbol{\Sigma}^2] = M\mathbf{I}_N$, and (78) comes from the results in (72) and (73). By further substituting (75), it yields

$$\log\mathbb{E}\left[\left\|\tilde{\mathbf{g}}_k^H\widehat{\mathbf{F}}\boldsymbol{\Sigma}\mathbf{V}^H\widehat{\mathbf{V}}\right\|^2\right] = \log M + \log\left(1 - \frac{M-N}{M}2^{-\frac{B_1}{M-1}}\right)$$
$$+ \log\left(\frac{1}{N} + \frac{N-1}{N}2^{-\frac{B_2}{N-1}}\right). \quad (79)$$

Then, by considering perfect feedback from relay to the BS, i.e., $\hat{\mathbf{v}}_j = \mathbf{v}_j$, we have another expectation result:

$$\mathbb{E}\left[\log\left(\left|\tilde{\mathbf{g}}_k^H\widehat{\mathbf{F}}\boldsymbol{\Sigma}\mathbf{V}^H\hat{\mathbf{v}}_j\right|^2\right)\right]$$
$$= \mathbb{E}\left[\log\left(\left|\tilde{\mathbf{g}}_k^H\widehat{\mathbf{F}}\boldsymbol{\Sigma}\mathbf{e}_j\right|^2\right)\right]$$
$$= \mathbb{E}\left[\log\left(\sigma_j^2\left|\tilde{\mathbf{g}}_k^H\hat{\mathbf{f}}_j\right|^2\right)\right]$$
$$= \mathbb{E}\left[\log\sigma_j^2\right] + \mathbb{E}\left[\log\left(\left|\tilde{\mathbf{g}}_k^H\hat{\mathbf{f}}_j\right|^2\right)\right]$$
$$= \frac{\log e}{N}\sum_{k=0}^{N-1}\psi(M-k) - \frac{\log e}{N-1}\sum_{k=1}^{2^{B_2}}\frac{1}{k} - \log e\sum_{k=1}^{N-2}\frac{1}{k} \quad (80)$$

where we use [28, Eq. 2.12]

$$\mathbb{E}\left[\log\sigma_j^2\right] = \frac{\log e}{N}\sum_{k=0}^{N-1}\psi(M-k) \quad (81)$$

and the expectation given in [20]

$$\mathbb{E}\left[\log\left|\tilde{\mathbf{g}}_k^H\hat{\mathbf{f}}_j\right|^2\right] = -\frac{\log e}{N-1}\sum_{k=1}^{2^{B_2}}\frac{1}{k} - \log e\sum_{k=1}^{N-2}\frac{1}{k}. \quad (82)$$

On the other hand, when $B_1$ is fixed and perfect CSI feedback is assumed from users to the relay, the above result becomes

$$\mathbb{E}\left[\log\left(\left|\tilde{\mathbf{g}}_k^H\widehat{\mathbf{F}}\boldsymbol{\Sigma}\mathbf{V}^H\hat{\mathbf{v}}_j\right|^2\right)\right]$$
$$= \mathbb{E}\left[\log\left(\sigma_k^2\left|\tilde{\mathbf{g}}_k^H\hat{\mathbf{f}}_k\right|^2\left|\mathbf{v}_k^H\hat{\mathbf{v}}_j\right|^2\right)\right]$$
$$= \mathbb{E}\left[\log\sigma_k^2\right] + \mathbb{E}\left[\log\left(\left|\tilde{\mathbf{g}}_k^H\hat{\mathbf{f}}_k\right|^2\right)\right] + \mathbb{E}\left[\log\left(\left|\mathbf{v}_k^H\hat{\mathbf{v}}_j\right|^2\right)\right]$$
$$= \frac{\log e}{N}\sum_{k=0}^{N-1}\psi(M-k) - \log e\sum_{k=1}^{N-1}\frac{1}{k} - \frac{\log e}{M-1}\sum_{k=1}^{2^{B_1}}\frac{1}{k}$$
$$- \log e\sum_{k=1}^{M-2}\frac{1}{k}. \quad (83)$$

Here, $\mathbb{E}[\log\sigma_k]$ is given by (81) and $\mathbb{E}[\log(|\tilde{\mathbf{g}}_k^H\hat{\mathbf{f}}_k|^2)]$ is from [22]

$$E\left[\log\left(\left|\tilde{\mathbf{g}}_k^H\hat{\mathbf{f}}_k\right|^2\right)\right] = \mathbb{E}\left[\log\text{beta}(1, N-1)\right] = -\log e\sum_{k=1}^{N-1}\frac{1}{k}. \quad (84)$$

The expectation of $\log(|\mathbf{v}_k^H\hat{\mathbf{v}}_j|^2)$ is obtained according to its distribution given by Lemma 1 in Appendix A.

## APPENDIX D
## SOLVING PROBLEM (47)

Letting $f(\theta) = \theta^m(1-\theta)^n$ with $m = M-1$ and $n = N-1$, problem (47) becomes

$$\underset{\theta}{\text{maximize}} \quad f(\theta)$$
$$\text{subject to:} \quad 0 < \theta < 1. \quad (85)$$

We compute

$$\frac{df(\theta)}{d\theta} = \theta^{m-1}(1-\theta)^{n-1}\left[m - (m+n)\theta\right]. \quad (86)$$

Setting the derivative to zero yields the unique stationary point of $f(\theta)$ over (0, 1) as

$$\theta^* = \frac{m}{m+n} = \frac{M-1}{M+N-2}. \quad (87)$$

The second-order derivative of $f(\theta)$ is found to be

$$\frac{d^2f(\theta)}{d\theta^2} = \theta^{m-2}(1-\theta)^{n-2}\left(-(m+n)\theta(1-\theta) + (m-(m+n)\theta)((m-1)-(m+n-2)\theta)\right) \quad (88)$$

that at the stationary point $\theta^*$ becomes

$$\left.\frac{d^2f(\theta)}{d\theta^2}\right|_{\theta=\theta^*} = -\left(\frac{m}{m+n}\right)^{m-2}\left(\frac{n}{m+n}\right)^{n-2}\frac{mn}{m+n} \quad (89)$$

which is strictly negative for any $M > 1$ and $N > 1$. Therefore, $\theta^*$ is the unique and global solution for problem (47).

## REFERENCES

[1] R. Pabst, B. H. Walke, and D. C. Schultz *et al.*, "Relay-based deployment concepts for wireless and mobile broadband radio," *IEEE Commun. Mag.*, vol. 42, no. 9, pp. 80–89, Sep. 2004.

[2] S. W. Peters, A. Y. Panah, K. T. Truong, and R. W. Heath, Jr., "Relay architectures for 3GPP LTE-advanced," *EURASIP J. Wirel. Commun. Netw.*, vol. 2009, 2009, Article ID 618787.

[3] S. Sesia, I. Toufik, and M. Baker, *LTE, the UMTS Long Term Evolution: From Theory to Practice*. New York: Wiley, Feb. 2009.

[4] B. Wang, J. Zhang, and A. Høst-Madsen, "On the capacity of MIMO relay channels," *IEEE Trans. Inf. Theory*, vol. 51, no. 1, pp. 29–43, Jan. 2005.

[5] X. Tang and Y. Hua, "Optimal design of non-regenerative MIMO wireless relays," *IEEE Trans. Wireless Commun.*, vol. 6, no. 4, pp. 1398–1407, Apr. 2007.

[6] O. Muñoz-Medina, J. Vidal, and A. Augstín, "Linear transceiver design in nonregenerative relays with channel state information," *IEEE Trans. Signal Process.*, vol. 55, no. 6, pp. 2593–2604, Jun. 2007.

[7] H. Bölcskei, R. U. Nabar, Ö. Oyman, and A. J. Paulraj, "Capacity scaling laws in MIMO relay networks," *IEEE Trans. Wireless Commun.*, vol. 5, no. 6, pp. 1433–1444, Jun. 2006.

[8] Y. Fan and J. Thompson, "MIMO configurations for relay channels: Theory and practice," *IEEE Trans. Wireless Commun.*, vol. 5, no. 5, pp. 1774–1786, May 2007.

[9] W. Yu and J. M. Cioffi, "Sum capacity of Gaussian vector broadcast channels," *IEEE Trans. Inf. Theory*, vol. 50, no. 9, pp. 1875–1892, Sep. 2004.

[10] N. Jindal, S. Vishwanath, and A. Goldsmith, "On the duality of Gaussian multiple-access and broadcast channels," *IEEE Trans. Inf. Theory*, vol. 50, no. 5, pp. 768–783, May 2004.

[11] H. Weingarten, Y. Steinberg, and S. Shamai, "The capacity region of the Gaussian MIMO broadcast channel," in *Proc. IEEE Int. Symp. Inf. Theory (ISIT)*, Chicago, IL, Jun. 2004, p. 174.

[12] T. Yoo and A. Goldsmith, "On the optimality of multiantenna broadcast scheduling using zero-forcing beamforming," *IEEE J. Sel. Areas Commun.*, vol. 24, no. 3, pp. 528–541, Mar. 2006.

[13] M. Sharif and B. Hassibi, "On the capacity of MIMO broadcast channels with partial side information," *IEEE Trans. Inf. Theory*, vol. 51, no. 2, pp. 506–522, Feb. 2005.

[14] R. Zhang, C. C. Chai, and Y. C. Liang, "Joint beamforming and power control for multiantenna relay broadcast channel with QoS constraints," *IEEE Trans. Signal Process.*, vol. 57, no. 2, pp. 726–737, Feb. 2009.

[15] W. Xu, X. Dong, and W.-S. Lu, "Joint optimization for source and relay precoding under multiuser MIMO downlink channels," in *Proc. IEEE ICC*, Cape Town, South Africa, May 2010, pp. 1–5.

[16] C. B. Chae, T. Tang, R. W. Heath, Jr., and S. Cho, "MIMO relaying with linear processing for multiuser transmission in fixed relay networks," *IEEE Trans. Signal Process.*, vol. 56, no. 2, pp. 727–738, Feb. 2008.

[17] F. A. Onat, H. Yanikomeroglu, and S. Periyalwar, "Relay-assisted spatial multiplexing in wireless fixed relay networks," in *Proc. IEEE GLOBECOM*, San Francisco, CA, Nov. 2006, pp. 1–6.

[18] K. Mukkavilli, A. Sabharwal, E. Erkip, and B. Aazhang, "On beamforming with finite rate feedback in multiple-antenna systems," *IEEE Trans. Inf. Theory*, vol. 49, no. 10, pp. 2562–2579, Oct. 2003.

[19] D. Love, R. Heath, and T. Strohmer, "Grassmannian beamforming for multiple-input multiple-output wireless systems," *IEEE Trans. Inf. Theory*, vol. 49, no. 10, pp. 2735–2747, Oct. 2003.

[20] N. Jindal, "MIMO broadcast channels with finite-rate feedback," *IEEE Trans. Inf. Theory*, vol. 52, no. 11, pp. 5045–5059, Nov. 2006.

[21] C. Au-Yeung and D. J. Love, "On the performance of random vector quantization limited feedback beamforming in a MISO system," *IEEE Trans. Wireless Commun.*, vol. 6, no. 2, pp. 458–462, Feb. 2007.

[22] T. Yoo, N. Jindal, and A. Goldsmith, "Multi-antenna downlink channels with limited feedback and user selection," *IEEE J. Sel. Areas Commun.*, vol. 25, no. 7, pp. 1478–1491, Sep. 2007.

[23] N. Ravindran and N. Jindal, "Limited feedback-based block diagonallization for the MIMO broadcast channel," *IEEE J. Sel. Areas Commun.*, vol. 26, no. 8, pp. 1473–1482, Oct. 2008.

[24] W. Xu, C. Zhao, and Z. Ding, "Optimisation of limited feedback design for heterogeneous users in MIMO downlinks," *IET Commun.*, vol. 3, no. 11, pp. 1724–1735, Nov. 2009.

[25] W. Dai, Y. Liu, B. Rider, and V. K. N. Lau, "On the information rate of MIMO systems with finite rate channel state feedback using beamforming and power on/off strategy," *IEEE Trans. Inf. Theory*, vol. 55, no. 11, pp. 5032–5047, Nov. 2009.

[26] B. Khoshnevis, W. Yu, and R. Adve, "Grassmannian beamforming for MIMO amplify-and-forward relaying," *IEEE J. Sel. Areas. Commun.*, vol. 26, no. 8, pp. 1397–1407, Oct. 2008.

[27] Y. Huang, L. Yang, M. Bengtsson, and B. Ottersten, "A limited feedback joint precoding for amplify-and-forward relaying," *IEEE Trans. Signal Process.*, vol. 58, no. 3, pp. 1347–1357, Mar. 2010.

[28] A. M. Tulino and S. Verdu, *Random Matrix Theory and Wireless Communications*. Hanover, MA: NOW Publishers, 2004.

[29] R. A. Horn and C. R. Johnson, *Matrix Analysis*. New York: Cambridge Univ. Press, 1985.

[30] I. S. Gradshteyn and I. M. Ryzhik, *Table of Integrals, Series, and Products*, 6th ed. San Diego, CA: Academic, 2000.

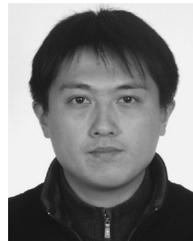

**Wei Xu** (S'07–M'09) received the B.Sc. degree in electrical engineering and the M.S. and Ph.D. degrees in communication and information engineering all from Southeast University, Nanjing, China, in 2003, 2006, and 2009, respectively.

He is currently an Associate Professor at the National Mobile Communications Research Laboratory (NCRL), Southeast University. Between 2009 and 2010, he was a Postdoctoral Fellow with the Department of Electrical and Computer Engineering, University of Victoria, Victoria, BC, Canada. His research interests include multi-antenna and multi-user channels, user scheduling, limited feedback strategies, and relay cooperative networks.

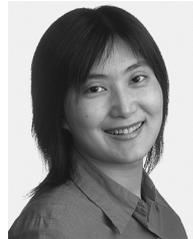

**Xiaodai Dong** (S'97–M'00–SM'09) received the B.Sc. degree in information and control engineering from Xi'an Jiaotong University, China, in 1992, the M.Sc. degree in electrical engineering from the National University of Singapore in 1995, and the Ph.D. degree in electrical and computer engineering from Queen's University, Kingston, ON, Canada, in 2000.

From 1999 to 2002, she was with Nortel Networks, Ottawa, ON, Canada, where she worked on the base transceiver design of the third-generation (3G) mobile communication systems. Between 2002 and 2004, she was an Assistant Professor at the Department of Electrical and Computer Engineering, University of Alberta, Edmonton, AB, Canada. She is presently an Associate Professor and Canada Research Chair (Tier II) at the Department of Electrical and Computer Engineering, University of Victoria, Victoria, BC, Canada. Her research interests include communication theory, cooperative communications and ultra-wideband radio.

Dr. Dong is an Editor for the IEEE TRANSACTIONS ON WIRELESS COMMUNICATIONS, the IEEE TRANSACTIONS ON VEHICULAR TECHNOLOGY, and the JOURNAL OF COMMUNICATIONS AND NETWORKS.

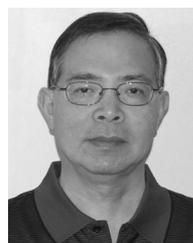

**Wu-Sheng Lu** (S'81–M'85–SM'90–F'99) received the B.Sc. degree in mathematics from Fudan University, Shanghai, China, in 1964 and the M.S. degree in electrical engineering and the Ph.D. degree in control science from the University of Minnesota, Minneapolis, in 1983 and 1984, respectively.

He was a Postdoctoral Fellow at the University of Victoria, Victoria, BC, Canada, in 1985 and a Visiting Assistant Professor with the University of Minnesota in 1986. Since 1987, he has been with the University of Victoria, where he is currently a Professor. His current teaching and research interests are in the general areas of digital signal processing and application of optimization methods. He is the coauthor with A. Antoniou of *Two-Dimensional Digital Filters* (Marcel Dekker, 1992) and *Practical Optimization—Algorithms and Engineering Applications* (Springer, 2007).

Dr. Lu is a Fellow of Engineering Institute of Canada.